# A model for hot spots and Q-slope in granular Niobium thin film superconducting RF cavities

Aymeric Ramiere[1], Claire Z. Antoine[2], and Jay Amrit[3, *]

[1]College of Physics and Optoelectronic Engineering, Shenzhen University, Shenzhen, 518060, China

[2]Université Paris-Saclay, CEA, Département des Accélérateurs, de la Cryogénie et du Magnétisme, 91191 Gif-sur-Yvette, France

[3]Université Paris-Saclay, CNRS, Laboratoire Interdisciplinaire des Sciences du Numérique, Rue du Belvédère, 91405 Orsay, France

* corresponding author: jay.amrit@universite-paris-saclay.fr

## Abstract

We propose a model to explain power dissipation leading to the formation of hot spots in the inner walls of niobium thin film superconducting RF cavities. The physical mechanism that we explore is due to the constriction of surface electrical current flow at grain interface boundaries. This constriction creates an additional electrical contact resistance which induces localized punctual heat dissipation. The temperature at these spots is derived; and the electrical contact resistance is shown to depend on the magnetic field, on the gain contact size over which dissipation occurs, and on other key parameters, including the effective London penetration depth and the frequency. The surface resistance and the quality factors are determined using our model and are shown to be in excellent agreement with experimental data.

# I. INTRODUCTION

Thin film superconductivity is an important pathway to improve the performance of superconducting RF (SRF) accelerators [1], [2]. Optimizing this technology requires a good understanding of the heat dissipation mechanisms. Niobium thin films grown mainly on copper substrates are deployed as a complementary to bulk niobium SRF cavities. Their performance at high fields is poor compared to bulk Nb, but they generally exhibit higher quality factors at low fields. The Nb films are mostly deposited by magnetron sputtering. SEM images from ref. [3] show that these films have a fiber-like structure composed of single-crystal grains with a typical diameter of 100 nm, that form a stack of vertical columns having grain boundaries mainly perpendicular to the surface. Closer to the substrate the grains are fine nanocrystalline single crystals, giving the film a porous granular aspect as the number of grain boundaries increases [3]. The porosity of the film increases at large sputtering tilt angles to the Nb surface [4]. Densification of the grain structure by high energy deposition techniques like HIPIMS has been shown to reduce porosity and partially mitigate the Q-slope [5], [6].

The thermal and electrical transport in granular media are limited by the grain boundaries when the mean free paths, $\ell$ of electrons and phonons become comparable to or greater than the size of grains. This is particularly true at low temperatures since the phonon density falls as $T^3$ whereas electrons in superconductors form Cooper pairs as the temperature decreases. The resistivity of grain boundaries is therefore affected by phonon and unpaired electron scattering at the boundaries. The arrangement and crystal orientation of grains also play a vital role in determining the grain boundary resistivity. These are some of the reasons why the characterization of the physical properties of grain boundaries is extremely complex. In addition, to our knowledge, there is no straightforward description of the topology of grain boundaries in that specific configuration as well.

In this paper, we propose a mechanism for heat dissipation in an attempt to explain the characteristic Q-slope behavior with the accelerating field in Nb thin film for SRF cavities. The mechanism we study is based on a well-known principle [7] that when an electrical current flows across two identical metals in contact, the current path is generally deviated and constricted depending on the nature of the contact before passing through a contact spot, as shown in Fig. 1 (a). The constriction of the current creates an additional electrical contact resistance in the vicinity of the boundary leading to heating. We conjecture that a similar phenomenon is possible at contacts between grain boundaries in a superconducting granular medium. The surface resistance expression in the vicinity of the dissipation site shall therefore be modified. In bulk niobium cavities, the large polycrystalline grains are densely packed and tightly connected by their grain boundaries, and the model, in its present form, hardly applies. In thin film, porosity and film densification are an issue, and the model is worth considering. In compliance with these ideas, we explore a new approach to explain the thermal behavior of Nb thin film SRF cavities for accelerators.

For simplicity, in our model we shall limit ourselves to ideal superconducting grains forming a lattice. The grain boundary structure between adjacent grains are considered to be all identical and each grain is connected to another through a square electrical conducting channel of size $a$ as shown in Fig. 1 (b). We also suppose that the grains are coupled together, and they form "tight links" since we consider the case where the grain boundary is superconducting. This model does not require any hypothesis on the shape of the cavity, but it is sensitive to the frequency of the RF field.

We shall firstly derive the equations for the temperature evolution and the power dissipated at a single grain boundary (SGB). Then we consider a lattice of grain boundary constrictions as shown in Fig. 1 (c). The temperature equation is then derived for a cluster of grain boundary constrictions. Using this equation, we compute numerically the temperature of the cluster as a function of the magnetic field for different cluster sizes. The dependence of the BCS surface resistance on the magnetic field and the electrical contact length, $a$ is put to evidence. The quality factors are calculated using our model and compared to experimental data available for Nb thin films.

## II. MODEL FOR DISSIPATION IN GRANULAR MEDIA

### A. Temperature at a single grain boundary (SGB)

Holm [7] defined the total electrical resistance between two identical materials as a "contact resistance" over a circular spot with a representative diameter $2a$ (often referred to as "*a- spot*" in the literature) given by $R_a = \frac{\rho}{\pi a} \arctan\left(\frac{z}{a}\right)$, where $\rho$ is the bulk resistivity of the metal and $z$ is the axis normal to the contact as shown in Fig. 1 (b). Associated with the current intensity $I$, there is a local dissipation of electrical power in the form of heat, due to the Joule effect at the boundary. The power at a single grain boundary can be written as $P_{sgb} = I^2 R_a(T_s) = U^2/R_a(T_s)$, where $R_a(T_s)$ is the electrical resistance associated to the temperature $T_s$ at the contact site and $U(T_s) = I R_a(T_s)$ is the local thermo-electrical potential which is established across the contact site. This phenomenon is present in many applications, for example in the automobile industry [8], thin films, porous copper [9] and metallic Josephson junctions [10].

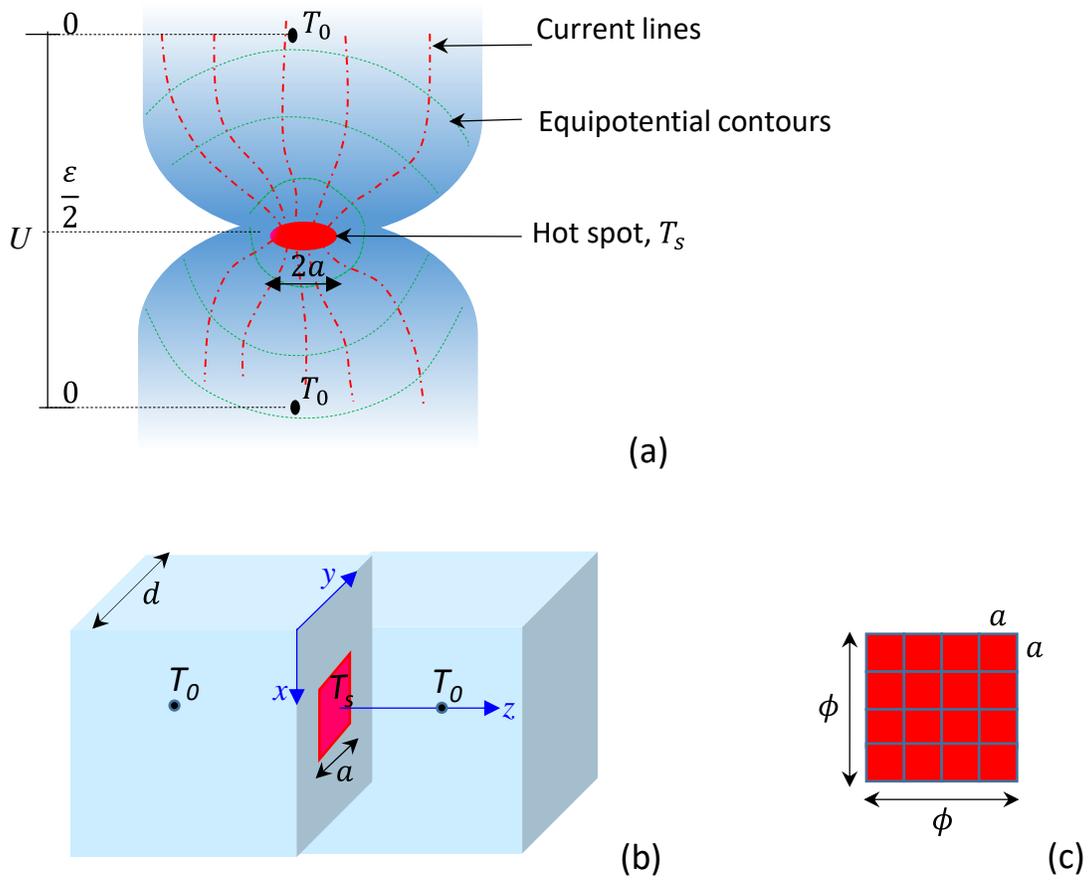

FIG. 1 (a). Holm model geometry showing of constriction of current lines between two grains, creating a hot spot at the contact boundary. The voltage drop $U(T)$ between $T_s$ and $T_0$ in each grain is equal to $\varepsilon/2$. (b). Schematics of our model representation of a square constriction of length $a$ between two grains of size $d$ in a Nb thin film. (c). Lattice of constriction sites forming a model square cluster of size $\phi$. (color online)

Electrical heat is dissipated at the constriction site, which becomes a hot spot. This heat is eventually carried away by thermal conduction cross-plane to the wall and into the system coolant (superfluid Helium in the case of superconducting cavities). In a straightforward 1D derivation of current flow across a single grain boundary, the local energy balance is written as:

$$-\kappa A \frac{dT}{dz} = I^2 R_a, \qquad (1)$$

where $T$ is the temperature, $\kappa$ is the thermal conductivity and $A$ is the cross-sectional contact area. This is a local equation and the heat dissipated depends on $R_a(T_s)$. The change in the electrical resistance $dR_a = \rho \frac{dz}{A} = \frac{dU}{I}$, where $\rho$ is the electrical resistivity due to normal conducting electrons, as explained later. Eq. (1) now becomes

$$-\kappa dT = \frac{I^2 R_a}{\rho} dR_a. \qquad (2)$$

Since the electrical resistivity and the thermal conductivity are due to electron flow, we can use the Wiedemann-Franz law $\kappa \rho = LT$, where $L$ is the Lorentz constant, and rewrite Eq. (2) as

$$-LT dT = U dU. \qquad (3)$$

Eq. (3) is in fact the Kohlrausch equation [7] which relates the temperature change to the voltage drop across metallic contacts. Integrating Eq. (3) with the boundary condition that the electrical potential is $U(T_0) = 0$ and $U(T_s) = \frac{\varepsilon}{2}$, we get

$$T_s^2 - T_0^2 = \frac{\varepsilon^2}{4L}. \qquad (4)$$

The magnitude of the local thermo-electrical potential, $\varepsilon$ is determined by taking the line integral of the surface electric field $E_y$ as seen by the electrons that "oscillate" at the hot spot, that is, $\varepsilon = \oint E_y \, dy = E_y \, a$, where $a$ corresponds to the length over which current constriction occurs at the grain boundary. From Eq. (4), the temperature of the hot spot is now given by

$$T_s = \left(T_0^2 + \frac{E_y^2 a^2}{4L}\right)^{\frac{1}{2}}. \tag{5}$$

The relationship between the induced surface electric field $E_y$, and the temperature of the hot spot becomes clearly highlighted. An important feature also emerging from Eq. (5) is the $T_s$ dependency on the size of the constriction $a$. We know that the electric field $E_y$ is induced by the time-varying magnetic field in the surface layer as shown by Faraday's law (see ref. [11]): $\oint E\, dx = \frac{d}{dt} \iint B\, dxdz$.

## B. Particular case: Superconducting material

In the case of Nb thin film SRF cavities, the screening current flows close to the inner surface of the cavity wall on a depth related to the field penetration depth $\lambda_L$. At T = 0 K, all the electrons from the conduction band are in the form of Cooper pairs. But at T > 0 K some pairs get broken by thermal activation. Because of the Cooper pairs inertia, the normal electrons issued from the broken Cooper pairs (quasiparticles) are sensitive to the RF field, and are at the origin of the surface resistance $R_{BCS}$ as formulated from the two-fluid model (see [13] and Eq 12b). Note that other sources of dissipation were not included in the initial description of the $R_{BCS}$ formula, and were attributed to a residual resistance term $R_{res}$. Lately, some efforts have been made to integrate other possible sources of pair breaking in the description [12] of the BCS surface resistance while external effects (like flux flow resistance) are kept in a residual term. Our model concerns normal conducting electrons and can be described as an additional possible origin of local heating. As described above, the constriction of the current may also lead to the formation of a hot spot. Faraday's law is also valid locally, and it therefore follows that:

$$E_y = -j\omega \lambda_L B_z, \tag{6}$$

where $\omega = 2\pi f$ is the angular frequency and $\lambda_L$ is the London penetration depth. Replacing in Eq. (5) now yields

$$T_s^2 = T_0^2 + \frac{(2\pi f)^2 \lambda_{eff}^2(T_s) B_z^2 a^2}{4L}. \tag{7}$$

The London penetration depth $\lambda_L(T_s) = \lambda_0/[1 - (T_s/T_c)^4]^{1/2}$ is replaced by the effective London penetration depth $\lambda_{eff}(T_s) = \lambda_L(T_s)\left(1 + \frac{\pi}{2}\frac{\xi_0}{\ell}\right)^{1/2}$, where $\xi_0 = 39$ nm for Nb, is the Cooper pair coherence length defined by the Pippard model [13], $\ell$ is the mean free path of the unpaired electrons and $\lambda_0 \approx 30$ nm. In this form the temperature equation reveals all the parameters that come into play in the formation of a hot spot in our model. As we shall see later, $\lambda_{eff}(T_s)$ plays a critical role in thermal runaway, especially as $T_s$ tends to the Nb transition temperature $T_c$.

Using equation (6), the power dissipated at a SGB, $P_{sgb} = \frac{\varepsilon^2}{R_a}$ simplifies to

$$P_{sgb} = \frac{(2\pi f)^2 \lambda_{eff}^2 B_z^2 a^2}{R_a}. \tag{8}$$

This equation shows that dissipation at a SGB depends on the frequency, the effective London penetration depth $\lambda_{eff}(T_s)$, the strength of the magnetic field, and the electrical contact length $a$ and resistance $R_a$. From Eq. (7), we have that the power dissipation has a quadratic dependence in temperature:

$$T_s^2 - T_0^2 = \frac{R_a}{4L} P_{sgb}. \tag{9}$$

The Lorentz constant plays a vital role in determining the temperature $T_s$. In most metals, it is generally equal to the theoretical value $L_0 = 2.45 \times 10^{-8}$ (V²/K²). For Nb in the superconducting state, it is shown that the Lorentz constant $L = 2.05 \times 10^{-8}$ (V²/K²) for $1.8 < T < 9.2$ K. [14] Firstly, considering the smallness of the Lorentz constant value, a minute increase in the electrical power dissipated across the contact will lead to a substantial increase in the hot spot temperature. Secondly, at sufficiently high currents, the temperatures at a grain boundary can increase very rapidly since $P_{sgb} \propto I^2$.

### C. Cluster of dissipating grain boundaries in a thin film

Typical Nb thin films deposited by magnetron sputtering exhibit small grains (typically 100 nm in diameter) with a fiber-like structure. The growth mechanism (inverted pyramids) induces some porosities and internal strain. The concentration of foreign atoms is high, originating from the tendency of Nb to react with light elements as well as the embedding of the discharge gas (usually Ar) [15]. Given the small size of grains, we expect that a misalignment of a single grain or the presence of impurities at a boundary shall affect the structure of neighboring grains within a certain radius. As a result, the current flow shall also be affected at different spots within this radius. We now consider such a cluster of $N_c$ grain boundary constrictions which, for simplicity, we shall assume forms a square lattice of dimension $\phi = a\sqrt{N_c}$ (see Fig. 1 (c)). Also, the grain boundary constrictions between adjacent grains are considered to be all identical and of length $a$ which simplifies the mathematical construction without loss of generality. In the realistic material, the "defective grain boundaries" are not necessarily regrouped into one single cluster. Here the cluster size $\phi$ is a convenient parameter that refers to the proportion of affected grain boundaries where current constriction occurs.

In the general treatment of multiple "spots" that are dispersed over a given radius, it is common practice for mathematical convenience to regroup the spots to form a cluster of size $\phi$ as for example in ref. [16]. We have considered the spots to be independent, that is, they conduct current in parallel. This is not necessarily a simplification as the electrical interaction between spots, if present, give negligible second-order modifications according to ref. [16].

The electrical power dissipated in the cluster of grain boundaries can be written as

$$P_m = P_{sgb} N_c, \qquad (10)$$

where $N_c = \frac{\phi^2}{a^2}$ is the number of electrical contact resistances and $P_{sgb}$ is given by Eq. (8). The temperature $T_m$ evolution of the cluster of SGB can now be expressed as

$$T_m^2 = T_0^2 + \frac{R_a}{4L} P_m \qquad (11a)$$

or more explicitly as

$$T_m^2 = T_0^2 + \frac{(2\pi f)^2 \lambda_{eff}^2 B_z^2 \phi^2}{4L}. \qquad (11b)$$

We suppose that the electron motion is such that the electron mean free path is predominantly limited by the smallest dimension, which corresponds to the length of the constriction channel, that is, $\ell \approx a$. Then, the temperature $T_m$ depends on the electrical contact length $a$ and the cluster size $\phi$. We note that Eq. (11b) must be solved iteratively since $\lambda_{eff}$ is a function of $T_m$.

### D. Electrical contact resistance at superconducting grain boundaries

In our model, the square electrical contact resistance between two grains as defined by Holm [7], simplifies to $R_a = \frac{\rho}{2a}$ when $z \geq a$. We now introduce the specific resistivity in units of $\Omega m^2$ as $\rho_s = \rho a = R_s a^2$, where the surface resistance of niobium is the sum of the BCS resistance and a residual resistance, that is, $R_s = R_{BCS} + R_{res}$. The contact resistance at the superconducting grain boundary can therefore be defined as

$$R_a(T) = R_s(T)/2, \tag{12a}$$

where the temperature is given by Eq. (11b). For the BCS resistance, we shall use the empirical equation [13], but the temperature is now replaced by $T_m$:

$$R_{BCS}(\Omega) = \frac{2 \times 10^{-4}}{T_m} \left(\frac{f}{1.5}\right)^2 exp\left(\frac{-17.67}{T_m}\right), \tag{12b}$$

where the frequency $f$ is in GHz. Eq. 12b is valid for temperatures up to $T_c$ [13]. From Eq. (12b) and Eq. (11b), we see that Eq. (12a) for the electrical contact resistance $R_a$ at the grain boundary acquires a complex dependency on the $\lambda_{eff}, a, B_z, f, L$ and $R_{res}$.

In our model $R_{res}$ is an adjustable parameter and its value must include resistances due to the presence of impurities at grain boundaries. The residual resistance $R_{res}$ varies from ~0 nΩ to some tens of nΩ for thin niobium films. The residual resistance also depends on other mechanisms such as the cooling history of the cavity, flux trapping and vortex dynamics.

## III. MODEL PREDICTIONS

To illustrate some of our model predictions, we need to set the frequency $f$ and the initial temperature $T_0$ of the inner cavity wall as the other input parameters in Eq. (11b).

## A. Evolution of the temperature of hot spots as a function of magnetic fields and electrical contact lengths

We calculate the temperature $T_m$ as the surface magnetic field $B_z$ is varied from 0 to 250 mT for hot spots of size $\phi$ formed by an agglomeration of identical grain constrictions. The frequency $f$ is set to 1.5 GHz and the inner wall initial temperature $T_0 = 1.7$ K so that comparison to experimental data is possible for the typical cases shown in Table I. The liquid helium operating temperature is $T_0$. Since $\lambda_{eff}$ depends on $T_m$, the calculations are therefore performed iteratively until both $T_m$ and $\lambda_{eff}(T_m)$ converge. Figure 2 shows examples for four different arbitrary cluster sizes, namely, 25, 40, 50, and 75 µm. Each of these clusters is composed of Nb grains having typical average electrical contact lengths of $a = 100$ nm.

The Fig. 2 shows that for $\phi \leq 25$ µm, the increase in $T_m$ with $B_z$ is noticeable, indicating the influence of power dissipation. Here, the magnetic fields can still reach an appreciable value (> 150 mT). However, for $\phi \geq 40$ µm the temperature $T_m$ increases rapidly and reaches the transition temperature, thereby limiting the maximum values of $B_z$ as can be observed from the dashed vertical lines inserted in the figure. The rate of change of $T_m$ with $B_z$ steepens to become exponential-like for these $\phi$ values. The dashed curve (violet) in Fig. 2 indicates the theoretical critical transition temperatures [17] for Nb at different magnetic fields. As is clear from Fig. 2, the quench is highly likely to occur at field values lower than the maximum attainable fields as the cluster sizes $\phi$ increases.

In thin films, a continuous decrease of the Q-factor (often referred as "Q slope") is generally observed over the whole range of accelerating fields, corresponding to a maximum magnetic field range of 60-150 mT. From Fig. 2 we can therefore identify that cluster sizes $\phi$ greater than ~40 µm play a predominant role in limiting the magnetic field. The functioning frequency of the cavity and the inner wall initial temperature $T_0$ also has a strong influence on the magnetic field. Increasing the frequency results in further limitations of the maximum attainable magnetic fields, $B_{max}$.

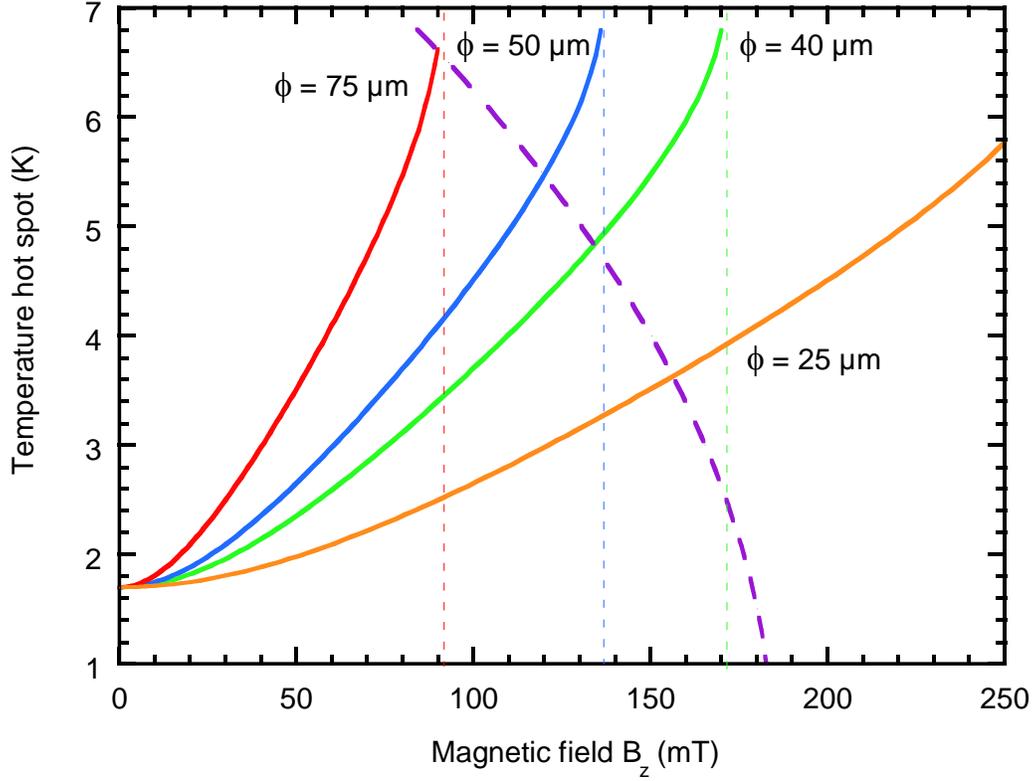

FIG. 2. Evolution of hot spot temperatures as the surface magnetic field is increased from 0 to 250 mT for different cluster sizes of 25, 40, 50 and 75 µm. Here, the electrical contact size is set to $a = 100$ nm for each cluster. The number of grain boundary constrictions $N_c$ increases with the cluster size. The vertical dashed lines highlight the shift of $B_{max}$ to lower fields as $\phi$ increases. The purple dashed curve corresponds to the theoretical transition from the Meissner state to the mixed state for pure niobium. It shows that early quench is to be expected even for the smaller clusters considered in this study.

Considering the grain electrical contact lengths $a$ in each cluster introduces an additional refinement to the results presented in Fig. 2. To examine this influence of the grain electrical contact length on the hot spot temperature, we plot in Fig. 3 this temperature as a function of the magnetic field $B_z$ for clusters, each of size $\phi = 50$ µm but having different electrical contact lengths. As before, the frequency is set to $f = 1.5$

GHz and the temperature of the inner wall $T_0 = 1.7$ K. The electrical contact lengths of 30 nm, 50 nm, 100 nm and 200 nm were considered, and the results are depicted in Fig. 3. The number of grain boundary constrictions, $N_c$ decreases as the lengths of the electrical contacts increase. The figure clearly shows that the maximum attainable magnetic field decreases as $a$ decreases, that is, as the number of grain electrical contacts increases. The figure also reveals that the magnetic field losses can reach ~ 30% depending on the electrical contact lengths in a given cluster. We note that at the maximum attainable field, the temperature of the cluster (hot spot) increases rapidly to approach the transition temperature $T_c$. From figures 2 and 3, it is obvious that large clusters of small grain electrical contact lengths lead to stronger limitations of the magnetic field.

Finally, we add that the cluster size $\phi$ in Eq. (11 b) is a convenient parameter that qualifies the effective length of the number of active SGB dissipation sites. These sites need not be grouped together but may be distributed throughout the cavity without altering the applicability of Eq. (11 b).

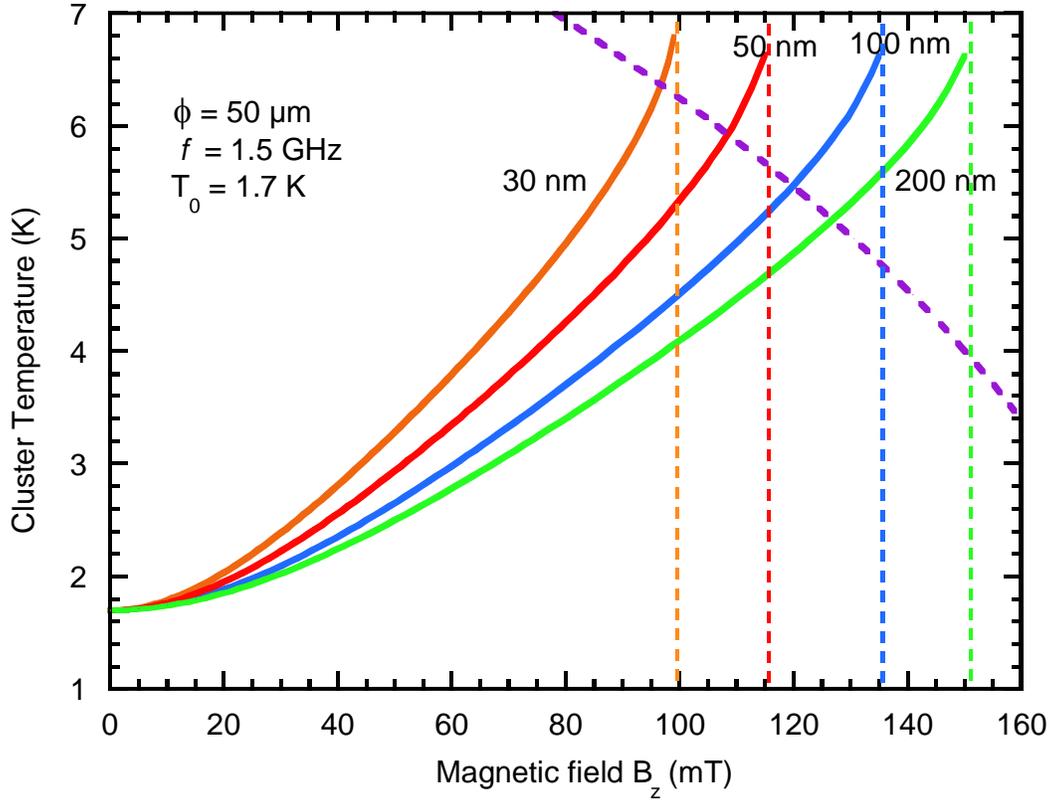

FIG. 3. The temperature of a cluster of electrical contacts at superconducting grain boundaries forming a hot spot of $\phi = 50$ µm. The orange, red, blue and green curves correspond respectively to electrical contact lengths of 30, 50, 100 and 200 nm. The purple dashed curve corresponds to the theoretical transition from the Meissner state to mixed the state for pure niobium.

### B. BCS surface resistance at a hot spot

The BCS surface resistance at a grain boundary, given by Eq. (12 b), acquires a complex dependency on the magnetic field and on the electrical contact size at grain boundaries, in addition to its dependency on temperature and frequency, as shown in Fig. 4. Here, the calculations are done at 1.5 GHz for three clusters of electrical contact lengths of 30, 50 and 100 nm respectively. At $B_z = 0$, the value of $R_{BCS}$ corresponds to its value in the absence of dissipation, with the initial temperature of the inner wall being $T_0 = 1.7$ K.

Fig. 4 shows that the $R_{BCS}$ increases rapidly with the magnetic field. Note that the temperature $T_m$ also increases with the magnetic field. At $T_m$ much greater than $T_0$, we have from Eq. (11 b) and Eq. (12 b) that $R_{BCS} \propto \frac{1}{\sqrt{N_c a}}$ to the first order. As is put to evidence in Fig. 4, the grain electrical contact lengths and the number of contact sites $N_c$ play an increasingly important role as the magnetic field increases.

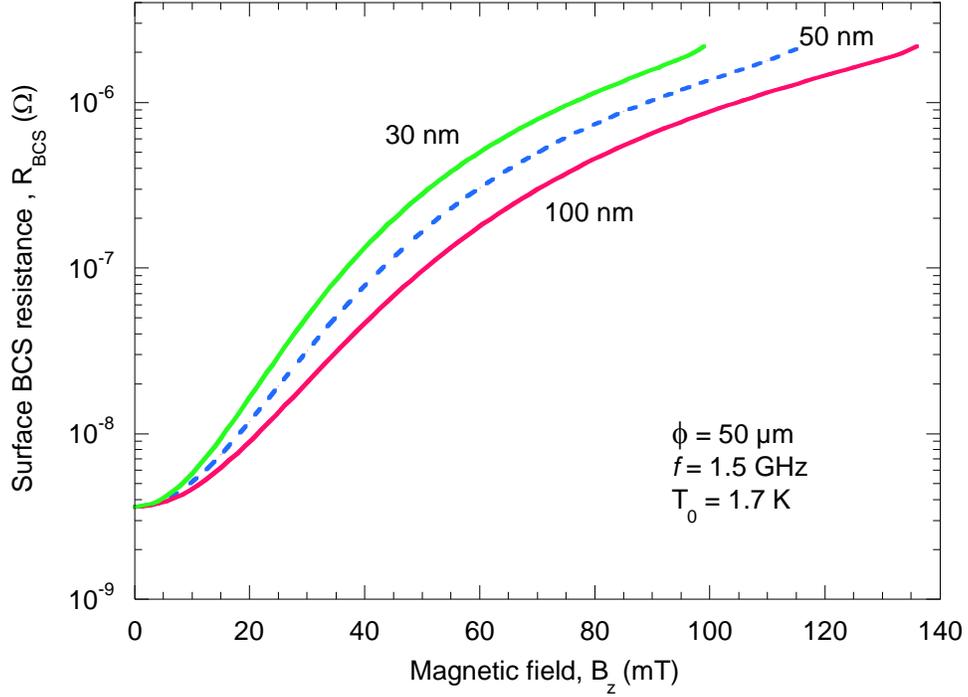

FIG. 4. The BCS surface resistance [Eq. (12 b)] as a function of $B_z$ for a cluster size of 50 µm and for three different grain electrical contact lengths, namely, 30 nm, 50 nm and 100 nm.

### C. Power dissipation in the inner cavity wall

Since the global power loss scheme in SRF cavities is due to Joule dissipation on the inner walls of the cavity when the magnetic field interacts with electrons, we calculate the power dissipated $\dot{q}$ (in W/m²) using the usual equation $\dot{q} = \frac{1}{2\mu_0^2} R_s(T_m) B_z^2$, where $\mu_0$ is the magnetic permeability of free space. Fig. 5 shows a plot of $\dot{q}$ as a function of the temperature $T_m$ for a cluster size $\phi = 50$ µm and for three different electrical contact lengths, namely, 30, 50 and 100 nm. The evolution of $T_m$ with $B_z$ is displayed in Fig. 3. The insert represents a zoom for $T_m < 4$ K. Actually, here the power corresponds to an upper limit since we have taken the inner wall temperature to be $T_m$. This is likely the case at high magnetic fields and therefore at high values of $T_m$ where the temperature spreads along the inner surface as it is conducted to the helium bath. We note that the copper substrate has a higher thermal conductivity than niobium. As shown in Fig. 4, the surface resistance $R_s(T_0)$ is one or two orders of magnitude smaller than $R_s(T_m)$; and therefore, the contribution of $R_s(T_0)$ to $\dot{q}$ is negligible. We also note that the orders of magnitude of the powers dissipated are in good agreement with our previous study given in ref. [18] where the energy balance is performed on the outer surface. Therefore, the impact of the Kapitza resistance at the Nb/superfluid interface and the thermal conductivity of niobium were considered in that study. In the present study we focus on a dissipative mechanism that influences Q-slope behavior in Nb thin films. Consequently, the Kapitza resistance, which is a second order effect in the present analysis, is neglected to avoid using additional adjustable parameters for clarity. Finally, our model reveals that the grain electrical contact lengths have a substantial effect on the power dissipated, especially at high magnetic fields as shown in Fig. 5.

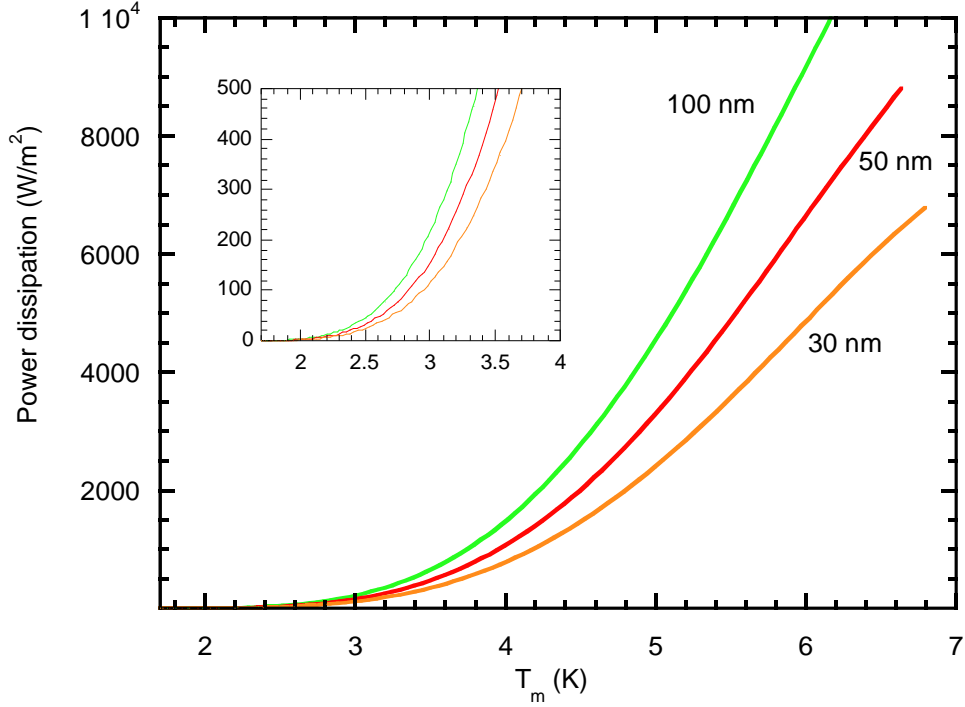

FIG.5 Power dissipation in three clusters, each 50 µm in size and having uniform grain electrical contact lengths of 30, 50 and 100 nm respectively. The insert is a zoom near the origin.

## IV COMPARISON WITH EXPERIMENTAL RESULTS

In the previous section we presented the general tendencies of $T_m(B_z)$, $R_s(B_z)$ and of power dissipation $\dot{q}(B_z)$ as predicted by our model. In this section we analyze different sets of experimental data using our model. Table I gives a summary of the three fitting parameters, namely the cluster size, the grain contact length and the residual resistance that come into play for the different experimental data sets taken into consideration. The justification of each parameter shall be found later in the text.

Table I: Fitting parameters for data of $Q_0(B_z)$ as shown in figures 7, 8, 9 and 10. $\phi$ is the cluster size, $a$ is the grain electrical contact length and $R_{res}$ is the residual resistance. For all data sets, the frequency is 1.5 GHz, except for data from ref. [19] where the frequency is 1.3 GHz.

|     | Data ref.     | $\phi$ (µm) | $a$ (nm)  | $R_{res}$(nΩ) | Figures     |
| --- | ------------- | ----------- | --------- | ------------- | ----------- |
| (a) | 1.7 K [20]    | 42 ± 2      | 60 ± 3    | 12 ± 2        | Fig.7 & 8   |
| (b) | 4.2 K [20]    | 60 ± 3      | 50 ± 5    | 12 ± 1        | Fig. 7      |
| (c) | 1.7 K [21]    | 15 ± 2      | 30 ± 2    | 8 ± 2         | Fig. 8      |
| (d) | 1.7 K [21]    | 18 ± 2      | 35 ± 2    | 0.01 ± 2      | Fig. 9      |
| (e) | 2.23 K [19]   | 40 ± 2      | 60 ± 3    | 0.1 ± 2       | Fig. 6 & 10 |

The error bars to our fitting parameters are estimated to encompass the scatter in the experimental data points. The error band surrounding each solid curve fit to the experimental dataset is represented by the shaded area, bounded by the maximum/minimum values of the fitting parameters. These bands indicate the sensitivity of the fitting parameters in our model.

## A. Surface resistance

In a study by Junginger [19] the field dependence of surface resistance of Nb/Cu superconducting cavities is measured at 1.3 GHz as a function of the accelerating field, $E_{acc}$. In Fig. 6, the blue dots represent one set of surface resistance measurements $R_s$ conducted at 2.23K (see Fig. 7 in ref. [19]). The magnetic field $B_z$ is determined using the relation $B_z(\text{mT}) = 4.26 E_{acc}(\text{MV/m})$ given in ref. [20]. We shall assume that $R_s = R_{BCS} + R_{res}$. The blue curve fitting the data is obtained using our model [Eq. (11 b) and Eq. (12 b)] with the cluster size $\phi = 40$ µm and the electrical contact lengths, $a = 60$ nm, as given in Table I (curve (e)). The number of electrical contacts $N_c \approx 4.4 \times 10^5$. As expected, $N_c$ is much smaller than the number of grains $N_g$ at the surface of a single mono-cell cavity having a typical surface of approximately 0.5 m². Here, $N_g$ is of the order of $5 \times 10^{13}$ grains/m² if one takes the average grain size to be of the order of 100 nm. Note that $R_{res}$ is assumed to be independent of the accelerating field in our fit and the best fit is obtained with $R_{res} = 0.1$ nΩ. For this set of data, the $R_{res}$ is completely negligible and $R_s$ is essentially due to $R_{BCS}$ as the magnetic field increases.

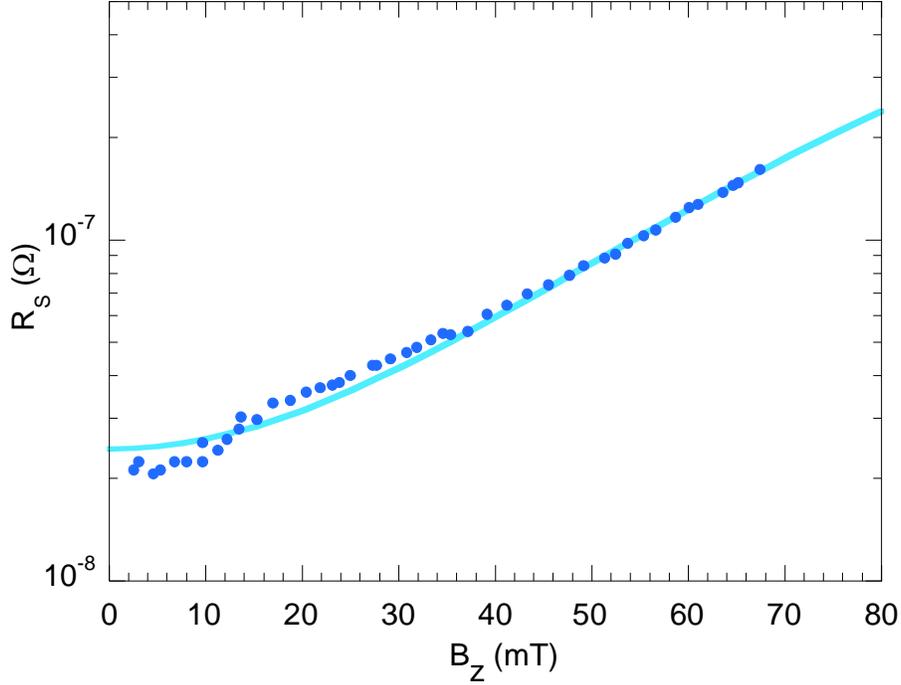

FIG. 6. The blue dots are surface resistance measurements of 1.3 GHz Nb/Cu cavity measured at 2.23 K, given in ref. [19] (data set (e) in Table I). The blue curve through data points is a fit obtained with our model at the same initial temperature.

## B. Q₀ curves fitting experimental data

The performance of a SRF cavity is characterized by the quality factor which can be defined as $Q_0 = G/R_s$, where $G$ is a constant that depends on the geometrical shape of the cavity and $R_s$ is the total surface resistance. We examine our model predictions of the quality factor in light of the experimental data available for thin films. The dependence of $R_{BCS}$ on the temperature $T_m$ given by Eq. 12 b, implies that the $Q_0$ in our model depends on the cluster size, electrical contact length and the magnetic field, through a "feedback" mechanism. The residual resistance $R_{res}$ has a direct impact on the initial $Q_0$ value. It is also important to note that $R_{res}$ is of the order of a few nanohms (see Table I) and it becomes completely negligible compared to $R_{BCS}$ as the magnetic field is increased (see Fig. 6). The value of $R_{res}$ therefore does not impact the values of the fitting parameters $\phi$ and $a$ found for the different curves.

Fig. 7 shows experimental data of the quality factor data from the work of Benvenuti *et.al.* [20] as a function of the magnetic field for thin Nb films grown on a copper substrate. The red dots correspond to measurements conducted at 1.7 K and the blue squares at 4.2 K. The measurements were performed at a frequency of 1.5 GHz on the same cavity. The curves going through the data were obtained using our model with different values of grain electrical contact lengths, cluster sizes and residual resistances given in Table I. The $R_{BCS}$ is determined using Eq. (12 b) which is valid at 1.7 K and at 4.2 K.

Lower (higher) values of the residual resistance have an effect of shifting the theoretical curve fitting the data vertically upwards (downwards) at the origin ($B_z = 0$) to higher (lower) $Q_0$ values. The cluster size tends to determine the slope of the $Q_0$ behavior at low fields. Smaller cluster sizes tend to flatten the slope of the $Q_0$ curve with $B_z$. The grain electrical contact lengths tend to have an effect on the slope of $Q_0$ at higher magnetic fields. Larger grain electrical contact lengths lead to more gradual slopes of $Q_0$ since power dissipation is reduced.

Our calculations of the $Q_0$ curves fit remarkably well the experimental data at both temperatures shown in Fig. 7. We used the value of $G = 295 \, \Omega$ and the relation $B(mT) = 4.55 E_{acc}$ given in ref. [ [20]]. The grain electrical contact lengths that are used in the fits are very plausible values given that the average grain sizes of $200 \pm 50$ nm were determined by TEM for these film structures. As expected, the residual resistance values found for each data set are in excellent agreement with each other and with the values given in that study [21].

Our model assumes that all grain boundaries have identical physical and geometrical properties. In reality, grain boundaries in a thin film are not all identical as they might differ in size, orientation, or impurity content… These latter elements can lead to thermal instabilities or percolation-like current flow during field increase, which may explain the slight discrepancies between our fits and the experimental data, as noticeable at low field values. The present model requires a better characterization of $R_{res}$ to account for this.

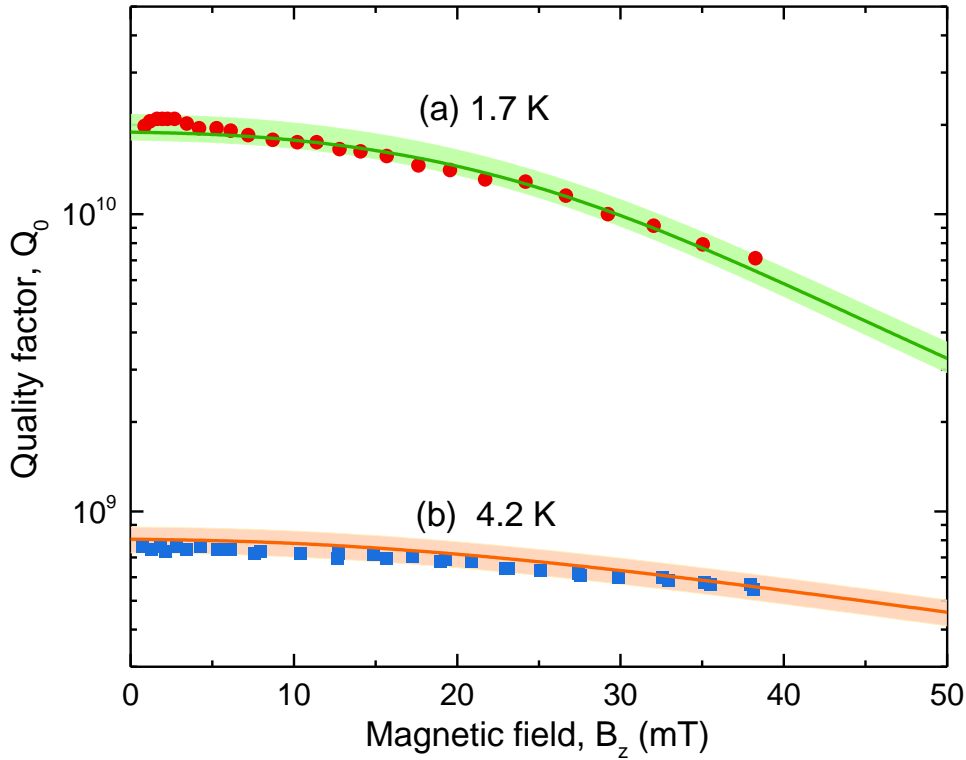

FIG 7. The red dots and blue squares correspond to quality factor data given in ref. [20] for temperatures 1.7 K and 4.2 K respectively. The green and orange curves are fits to these data using the parameters values given Table I. The shaded areas are obtained with the error bars given in Table I.

The blue squares in Fig. 8 are data at 1.7 K, which is also measured on a thin film Nb deposited on electropolished Cu substrate given in ref. [21]. The characteristics of the curve (c) fit shown in this figure are given in Table I. For the sake of comparison, we replotted here the data set at 1.7 K shown in Fig. 7 (curve (a)). Both sets of data in Fig. 8 were conducted at $f = 1.5$ GHz. The discrepancy between the two curves (a) and (c), which fit remarkably well both sets of the experimental data, is due to the cluster size, the inter-grain electrical contact lengths and the residual resistance, as shown in Table I. As discussed above, dissipation occurs at smaller cluster sizes for the blue square data and therefore lead to higher $Q_0$ value at higher magnetic fields compared to red dot data (a). From Table I, the number of electrical contacts $N_c$ found for curve (a) is $\sim 5 \times 10^5$ whereas it is $\sim 2.5 \times 10^5$ for the curve (c). The results, therefore, suggest that the $Q_0$ value is predominately controlled by $N_c$. We also note that the shift in the initial $Q_0$ value is attributed to the residual resistance which clearly plays an important role in defining the maximum $Q_0$ values at $B_z \approx 0$.

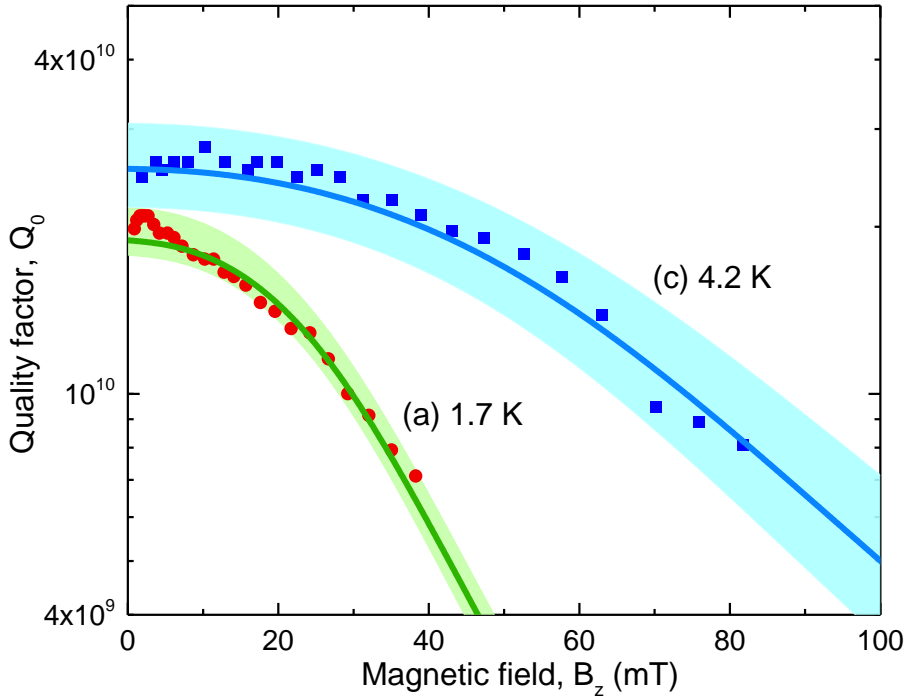

FIG.8. Two sets of data at 1.7 K and at 1.5 GHz. The blue squares are data given in ref. [21] and the red dots are from ref. [20]. Both curves fitting the data sets are determined using our model.

Fig. 9 shows another set of data (red dots) also taken on thin film Nb on Cu substrate given in ref. [21]. The details pertaining to the curve (d) obtained using our model are given in Table I. The cluster size and grain electrical contact lengths are slightly larger than the characteristics of the curve (c) in Fig. 8. This is so because at high magnetic fields, the rate of change of $Q_0$ with the magnetic field is bigger here than for the data in Fig. 8. This implies stronger dissipation due to larger cluster sizes as indicated by the values given in Table I. Our model fit is in very good agreement with the data points at fields greater than 40 mT. At lower fields, our curve lies within the error bars associated with the data as given in ref. [21].

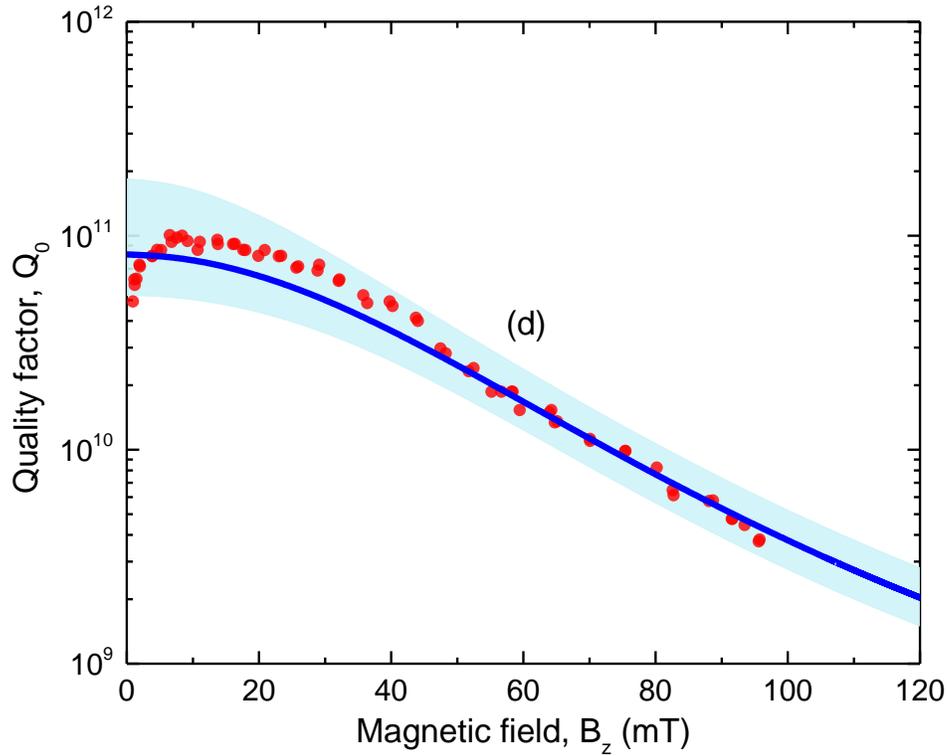

FIG. 9. Red dots correspond to data taken from ref. [21]. The blue curve is a fit using our model. The cluster size, electrical contact lengths and residual resistance determined from the fit are given in Table I. The error bars given in Table I define the shaded area.

Using the experimental data [19] of $R_s$ represented in Fig. 6, we show (blue dots) in Fig. 10 the quality factor as a function of the magnetic field for the Nb/Cu cavity at 2.23 K. Here, the quality factor is determined using $Q_0 = G/R_s$, with $G = 270$. The red curve is obtained from our model with the values of the adjustable parameters given in Table I [curve e)]. As in all the data analyzed above, there is a very good agreement between experiments and our model, especially above ~10 mT.

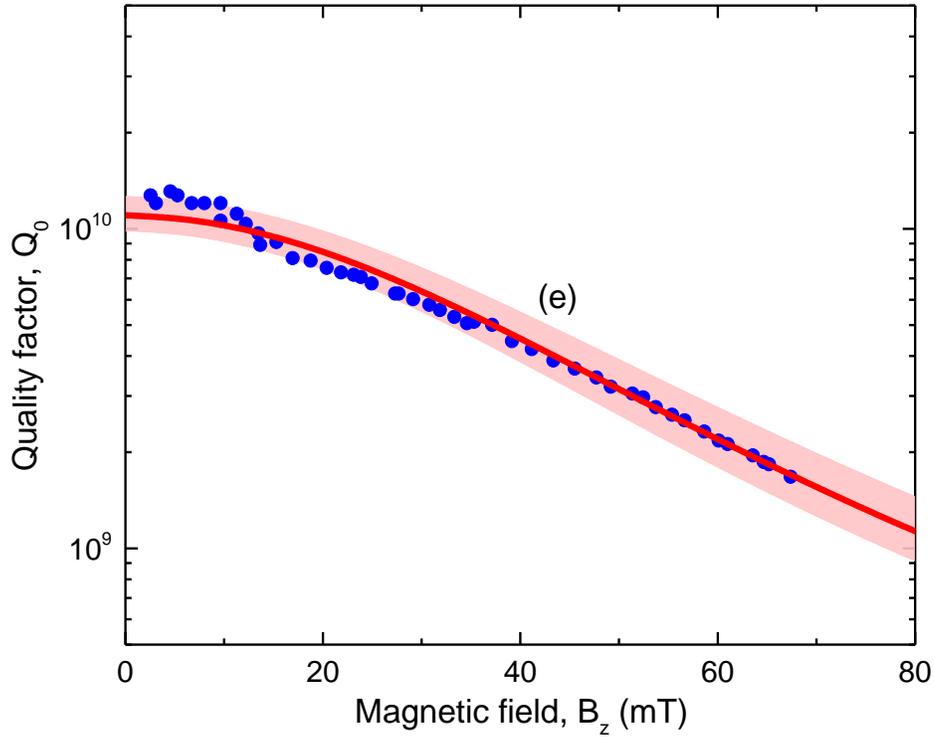

FIG. 10. Blue dots are experimental values at 2.23 K of the quality factor as a function of the magnetic field for a Nb/Cu cavity functioning at 1.3 GHz (see ref. [19]). The red curve is a fit determined using our model. Refer to Table I for the fitting parameters.

We also note that for Figs. 9 and 10 the error bar estimations of $R_{res}$ given in Table I are large. This is due to the origin of the experimental error bars at low magnetic fields. Indeed, usually, the most accurate measurements of $Q_0$ are achieved near critical coupling. Fixed length antennas are installed in cavities such that the measurements are the most accurate at high magnetic fields so as to determine their best performance. Inevitably, the errors bars are higher at low magnetic fields, as indicated in ref. [22] and especially for the datasets represented in Figs. 9 and 10 where the $Q_O$ values are very high around $B_z = 0$.

In Fig. 11 we show the temperature increase $\Delta T = (T_m - T_0)$ at a hot spot as the $Q_0$-values evolve with the increase in the magnetic field, $B_z$. The figure is plotted for data (d) shown in Table I and displayed in Fig. 9. For clarity we show the $B_z$ on the left y-axis. The figure indicates orders of magnitude of temperature differences that come into play for this data set. For example, at the highest field $B_z = 90$ mT, the temperature increase $\Delta T$ is approximately 0.8 K. Finally, we note that these orders of magnitude of $\Delta T$ are difficult to measure. Indeed, unlike bulk Nb cavities, the high thermal conductivity of copper allows heat to spread laterally within the cavity wall and thereby compromises the detection of localized hot spots on the outer-side of the cavity wall.

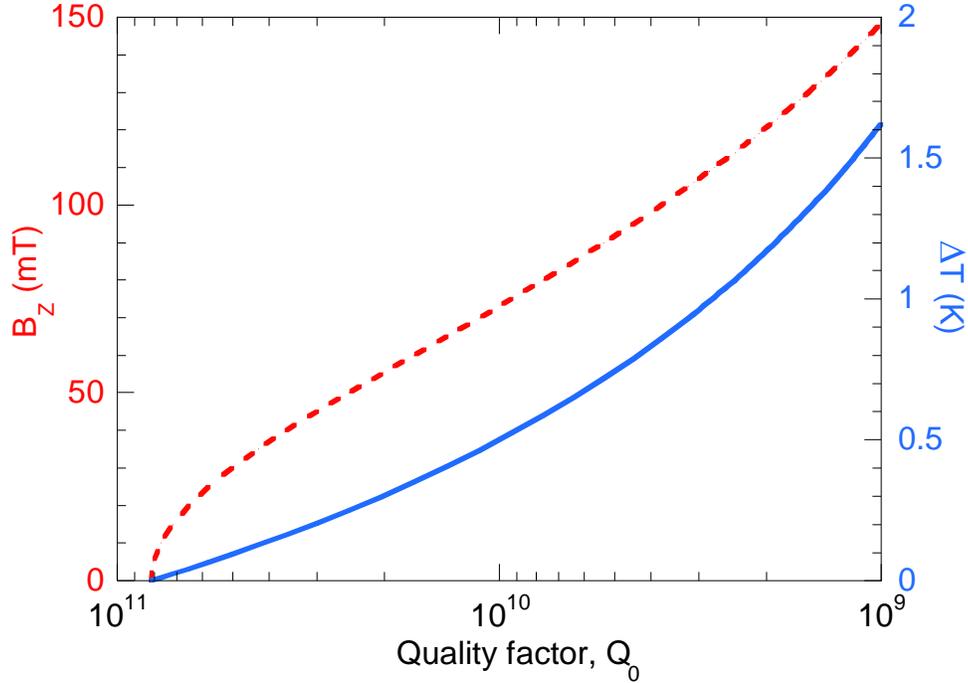

FIG. 11. Evolution of the temperature increase of a hot spot from its initial value $T_0 = 1.7$ K as the quality factor decreases with the increase in the magnetic field. The red dashed curve corresponds to the magnetic field and the blue curve to the temperature increase. The plot is done for data set (d) in Table I.

## V. DISCUSSIONS

Our model reveals that the Q slope observed on thin film cavities could be attributed to an important source for Ohmic heating due to the constriction of surface current flux in the inner cavity wall. Since the contact resistance, $R_a$ is strongly related to the grain electrical contact length, $a$, it follows that the contact resistances can be modified by tailoring the growth of the Nb thin film. For example, the HiPIMS fabrication technique of thin films provides denser columnar material, which should help increase the contact lengths and therefore decrease the number of contact resistances.

Dissipation in granular Nb superconducting films has been examined in various studies. [23], [24], [25] In these models, the grains are linked to each other by boundaries. The magnetic field penetration at grain boundaries is studied by taking into account the grain size and by considering whether the coupling of grains is classified as "weak links" or "strong links". The latter is determined by the critical current density $J_c$ and the specific resistance. [23] In these studies, the grain and the grain boundaries are modeled as an electrical RL circuit. In the present model we have adopted a thermal approach and we have derived the temperature at a hot spot. We note that in all studies, including the present, the surface resistance depends on the electrical properties at the superconducting grain boundaries.

In our model, the BCS surface resistance [Eq. (12 b)] depends on the grain electrical contact length, $a$ rather than the grain size $d$ as in the previous model. [23] These two parameters are indeed related since the grain electrical contact length is always less than or equal to the grain size. Firstly, we also note that in our model, the constriction length associated with a cluster is an average over the $N_c$ electrical contact lengths present. A cavity can have a distribution of grain electrical contact lengths. Further refinement is required in our model to take into account this aspect. Secondly, since current conduction is dependent on grain orientation, grain orientation anisotropy must therefore influence the surface resistance. [25] In the present study, the increase in the surface resistance is a direct consequence of an anisotropic current flow which can also be due to misoriented grains, in addition to the presence of impurities in the cavity wall. Thirdly, our study tends to show that the $R_s$ is dominated by $R_{BCS}$ as the fields increase; and $R_{BCS}$ turns out to be one or two orders of magnitude greater than $R_{res}$.

# VI. CONCLUSIONS

We have investigated heat dissipation in niobium thin film SRF cavities to explain the Q-slope behavior with the accelerating fields. In our model, the dissipative mechanism is due to the constriction of surface current flow that creates electrical contact resistances at grain boundaries. The temperature increases at a single grain boundary [Eq. (7)] and at a cluster of single grain boundary constrictions [Eq. (11 b)] are derived for the first time and solved iteratively. Our model reveals the relevant parameters that come into play, including the frequency, the effective London penetration depth, the grain electrical contact length, the cluster size and therefore the number of electrical contact resistances, the Lorentz constant and the mean free path of electrons. The BCS resistance in the vicinity of the grain boundary is shown [see Fig. 4] to acquire a complex dependency on these parameters. As demonstrated, our model gives an excellent prediction of the experimental Q-slope measurements in different studies, using the residual resistance, the electrical contact length and cluster size as fitting parameters and therefore provides a quantitative understanding of the Q-slope phenomenon in Nb thin film SRF cavities. The orders of magnitude of the electrical contacts lengths lie between 30-100 nm and the number of electrical contacts is of the order of $10^5$. We believe that the present model can be used as a useful tool to analyze the Q-performance of Nb superconducting thin film cavities.

# Acknowledgments

We express our thanks to members of the Irene Joliot-Curie Laboratory (IJC Lab) in Orsay, especially G. Martinet, M. Fouaidy and D. Longuevergne for discussions.